\documentclass[fleqn,10pt]{wlscirep}

\usepackage{graphicx}
\usepackage{amsmath}
\usepackage{bm}
\usepackage[normalem]{ulem}
\usepackage{caption}

\title{Vehicles as sensors: high-accuracy rainfall maps from windshield wiper measurements}

\author[1,+]{Matthew Bartos}
\author[2,+]{Hyongju Park}
\author[2]{Tian Zhou}
\author[1,*]{Branko Kerkez}
\author[2]{Ramanarayan Vasudevan}
\affil[1]{Department of Civil and Environmental Engineering,
  University of Michigan, Ann Arbor, MI, 48109, United States}
\affil[2]{Department of Mechanical Engineering,
  University of Michigan, Ann Arbor, MI, 48109, United States}

\affil[*]{Correspondence to: bkerkez@umich.edu}

\affil[+]{These authors contributed equally to this work}

\begin{abstract}
  Connected vehicles are poised to transform the field of environmental sensing
  by enabling acquisition of scientific data at unprecedented scales. Drawing on
  a real-world dataset collected from almost 70 connected vehicles, this study
  generates improved rainfall estimates by combining weather radar with
  windshield wiper observations. Existing methods for measuring precipitation
  are subject to spatial and temporal uncertainties that compromise
  high-precision applications like flash flood forecasting. Windshield wiper
  measurements from connected vehicles correct these uncertainties by providing
  precise information about the timing and location of rainfall. Using
  co-located vehicle dashboard camera footage, we find that wiper measurements
  are a stronger predictor of binary rainfall state than traditional stationary
  gages or radar-based measurements. We introduce a Bayesian filtering framework
  that generates improved rainfall estimates by updating radar rainfall fields
  with windshield wiper observations. We find that the resulting rainfall field
  estimate captures rainfall events that would otherwise be missed by
  conventional measurements. We discuss how these enhanced rainfall maps can be
  used to improve flood warnings and facilitate real-time operation of
  stormwater infrastructure.
\end{abstract}
\begin{document}

\flushbottom
\maketitle
\thispagestyle{empty}

\section*{Introduction}

Accurate rainfall measurements are essential for the effective management of
water resources \cite{Overeem_2013}. Historical rainfall records are used
extensively in the design of water infrastructure \cite{Cheng_2014}, while at
finer scales, real-time rainfall measurements are an integral component of flood
forecasting systems \cite{Hapuarachchi_2011}. Despite the central role that
precipitation measurements play in the design and operation of water
infrastructure, current methods for measuring precipitation often do not provide
the spatial resolution or measurement certainty required for real-time
applications \cite{Hapuarachchi_2011}. As the demand for real-time precipitation
data increases, new sensing modalities are needed to address deficiencies found
in conventional data sources.

The need for high-resolution precipitation estimates is perhaps best illustrated
by the problem of urban flash flooding. Flooding is the number one cause of
natural disaster fatalities worldwide, with flash floods accounting for a
majority of flooding deaths in developed countries \cite{Doocy_2013}. Despite
the risks posed by flash flooding, there is ``no existing model [that is]
capable of making reliable flash flood forecasts in urban watersheds''
\cite{Hapuarachchi_2011}. Flash flood forecasting is to a large extent hindered
by a lack of high-resolution precipitation data, with spatial resolutions of $<$
500 m and temporal resolutions of 1-15 minutes required for urban areas
\cite{Berne_2004, Smith_2007}.

\begin{figure*}[!htb]
\centering
\includegraphics[width=\textwidth]{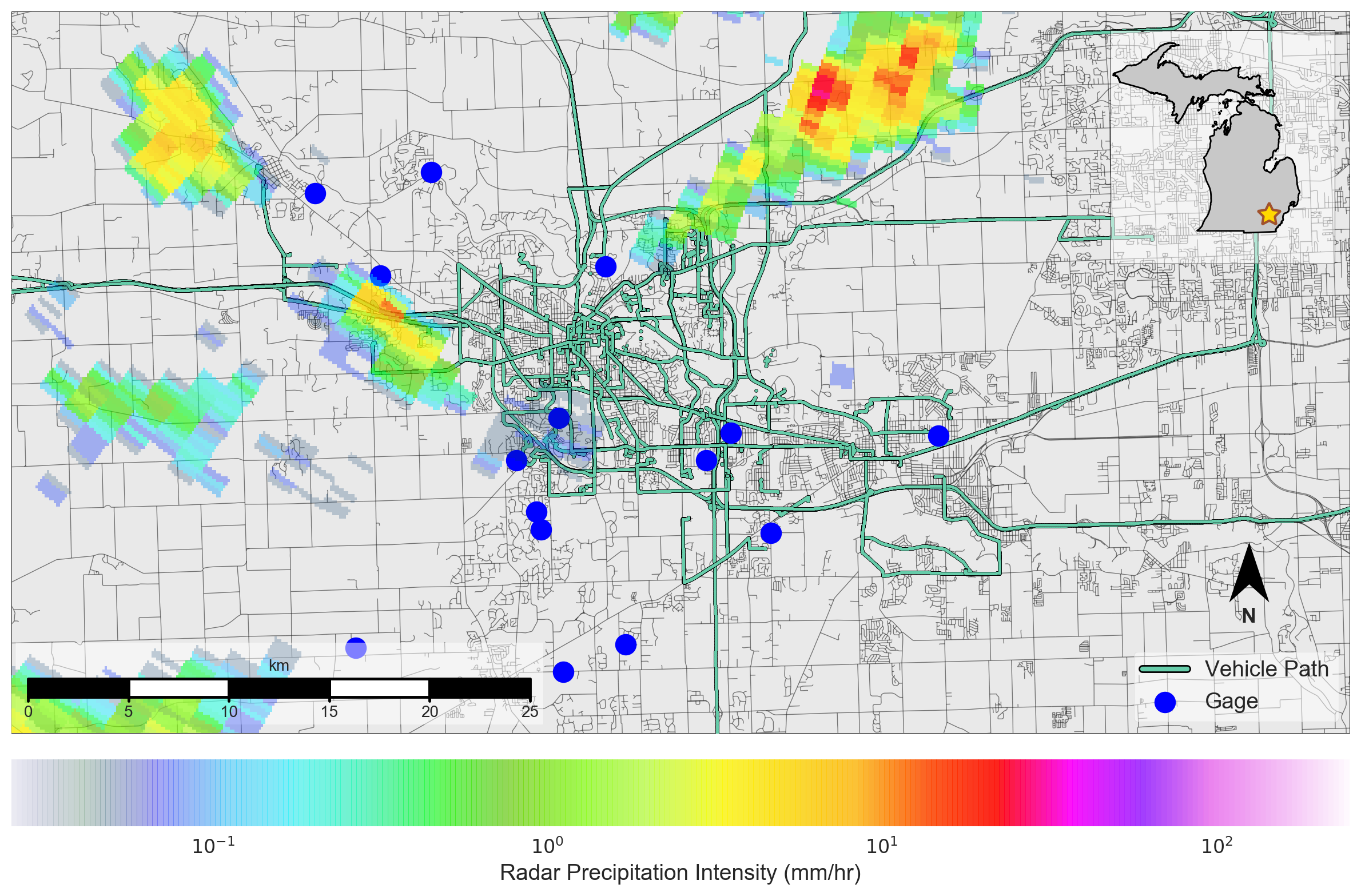}
\caption{ \textbf{Overview of the study area on June 12, 2014.} Blue circles
  represent rain gages. Vehicle paths are shown as green lines, while roads are
  shown in gray. A radar overlay shows the average precipitation intensity as
  estimated by radar.}
\label{overview_map}
\end{figure*}

Contemporary rain measurement technologies---such as stationary rain gages and
weather radar---struggle to achieve the level of precision necessary for flash
flood forecasting. While rain gages have long served as a trusted source of
surface-level precipitation measurements \cite{Grimes_1999}, they often fail to
capture the spatial variability of rain events, especially during convective
storms \cite{Xiaoyang_2003, Yilmaz_2005, Sun_2000}. This inability to resolve
spatial patterns in rainfall is made worse by the fact that the number of rain
gages worldwide is rapidly declining \cite{Overeem_2013}. Weather radar is a
useful tool for capturing the spatial distribution of rainfall. However,
radar-rainfall estimates are subject to large spatial and temporal uncertainties
\cite{Winchell_1998, Morin_2003, Smith_1996, Islam_2012}. Additionally, weather
radar tends to show systematically large biases for major flood events, and may
perform poorly for small watersheds \cite{Smith_2007}, making urban flood
forecasting problematic.

The rise of connected and autonomous vehicles offers an unprecedented
opportunity to enhance the density of environmental measurements
\cite{Hill_2015, Haberlandt_2010}. While dedicated sensor networks are expensive
to deploy and maintain, fleets of connected vehicles can capture real-time data
at fine spatial and temporal scales through the use of incidental onboard
sensors. With regard to rainfall measurement, windshield wiper activity offers a
novel means to detect the location and timing of rainfall with enhanced
precision. When used in conjunction with modern signal processing techniques,
wiper-based sensing offers several attractive properties: (i) vehicles achieve
vastly improved coverage of urban areas, where flood monitoring is important;
(ii) windshield wiper intensity is easy to measure and requires little overhead
for processing (as opposed to video or audio data); and (iii) vehicle-based
sensing can be readily scaled as vehicle-to-infrastructure communication becomes
more widespread. Moreover, many new vehicles come equipped with optical rain
sensors that enable direct measurement of rainfall intensities. When paired with
data assimilation techniques, these sensors may enable even
higher-accuracy estimation of rainfall fields compared to wipers alone.

While a small number of studies have investigated vehicle-based precipitation
measurements, the results of these studies are strictly based on simulated wiper
data instead of real measurements. As such, the premise that windshield wiper
data can be used to improve rainfall estimates has never been verified using a
large real-world dataset. Hill (2015) combines simulated binary (wet/dry)
rainfall sensors with weather radar observations to generate improved areal
rainfall estimates, which are then validated against rainfall fields produced by
interpolation of tipping-bucket rain gages \cite{Hill_2015}. Similarly,
Haberlandt (2010) combines simulated vehicle wiper measurements with rain gage
observations to improve rainfall field estimates, and then validates the
resulting product against weather radar \cite{Haberlandt_2010}. Although these
studies highlight the potential for vehicle-based measurements to improve the
spatial and temporal resolution of rainfall estimates, their findings have not
yet been validated using data from real-world connected vehicles.

To address these challenges, this study leverages windshield wiper measurements
collected from nearly 70 vehicles to produce corrected rainfall maps (see
Figure \ref{overview_map} for a description of the study area and data sources). In the
first part of this paper, we demonstrate that windshield wiper measurements
offer a reliable indicator of rainfall by comparing wiper measurements against
dashboard camera footage that indicates the ground truth binary rainfall state
(raining/not raining). In the second part of this paper, we develop a Bayesian
data fusion procedure that combines weather radar with vehicle-based wiper
measurements to produce an updated probabilistic rainfall field map. We validate
this novel data product by showing that it is more effective than the original
radar data at predicting the binary rainfall state. Finally, we discuss how
these enhanced rainfall maps can be used to improve flood warnings and
facilitate real-time operation of stormwater infrastructure.

\section*{Results}

\begin{figure*}[!htb]
\centering
 \includegraphics[width=\textwidth]{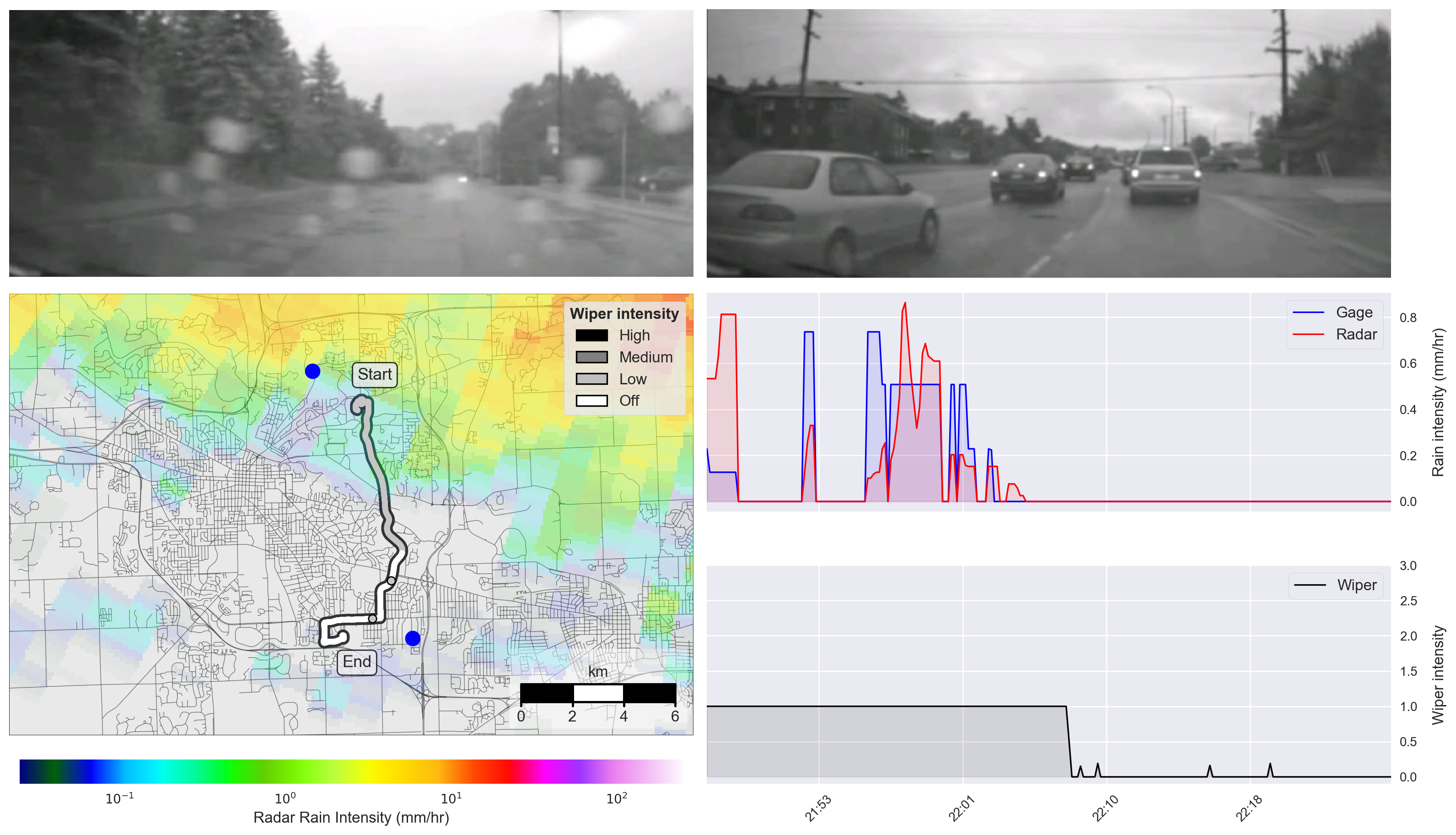}
 \caption{ \textbf{Analysis of a single vehicle trip occurring from 21:46 - 22:26 on
   August 11, 2014.} The top two panels show video footage during the
   rainy (left) and dry (right) segments of the trip. The bottom left panel
   shows a map of the vehicle's trip, with the wiper intensity indicated by
   color. A radar overlay shows the average rainfall intensity over the
   40-minute time period. Blue circles represent the gages nearest to the
   vehicle path. The two bottom right panels show the precipitation intensity as
   estimated by radar and gage measurements (center), and the 1-minute average
   wiper intensity (bottom).}
\label{trip_analysis}
\end{figure*}

\subsection*{Windshield wipers improve binary rainfall detection}

Windshield wiper measurements enhance rainfall estimation by enabling greater
certainty about the timing and location of rainfall. While wiper intensity is
generally a poor predictor of rainfall intensity (see Figure S1 in the
Supplementary Information), we find that wiper status (on/off) is a stronger
predictor of binary rainfall state than either radar or gage-based measurements.
This result suggests that vehicle-based measurements can be used to validate and
correct rainfall fields derived from conventional data sources.

Wiper measurements provide a more accurate indicator of binary rainfall state
than either radar or gage measurements. We determine the binary classification
performance for each technology (gages, radar and wipers) by comparing the
measured rainfall state with co-located dashboard video footage. Dashboard video
is taken to represent the ground truth, given that the presence or absence of
rainfall can readily be determined by visually inspecting the windshield for
raindrops. Figure \ref{trip_analysis} shows an example of co-located radar,
gage, wiper and camera measurements for a single vehicle trip. In this case,
rainfall is visible during the first half of the trip (top left). Aggregating
these cross-comparisons for every vehicle trip across three storm events, we
find that wiper status is the best estimator of binary rainfall state, with a
true positive rate (TPR) of 93.1\%, and a true negative rate (TNR) of 98.2\%. By
comparison, weather radar achieves a smaller TPR of 89.5\%, while stationary
gages show a much smaller TPR of 44.5\% (see Table \ref{tab:tab1}). These
results can partly be explained by the superior spatial and temporal resolution
of the vehicle-based measurements. Wipers detect intermittent changes in
rainfall at a temporal resolution on the order of seconds, while radar and gage
measurements can only detect the average rate over a 5-minute period. When
ground truth camera observations are collected at a 3-second temporal
resolution, the benefit of wiper measurements over radar measurements becomes
even more pronounced, with a TPR advantage of 5.2\%, a TNR advantage of 7.7\%,
and an overall wiper TPR of 97.0\% (see the supplementary note on factors
affecting binary detection performance). The results of this analysis suggest
that conventional rainfall measurement technologies can be enhanced through the
inclusion of vehicle-based measurements.

\begin{table}[!htb]
  \centering
  \begin{tabular}{| l | l l l |}
    \hline
    Metric & Gage & Radar & Wiper \\
    \hline
    True Positive Rate (\%) & 44.5 & 89.5 & 93.1 \\
    True Negative Rate (\%) & 96.7 & 97.5 & 98.2 \\
    \hline
  \end{tabular}
  \caption{\textbf{Classification performance of each rainfall measurement
      technology.} The true positive rate indicates the percentage of instances
    where the given technology successfully detects rainfall when rainfall is
    actually occurring. The true negative rate indicates the percentage of
    instances where the technology does not detect rainfall when rainfall is not
    occurring.}
  \label{tab:tab1}
\end{table}

\subsection*{Assimilation of wiper data yields corrected rainfall maps}

\begin{figure*}[!htb]
\centering
\includegraphics[width=\textwidth]{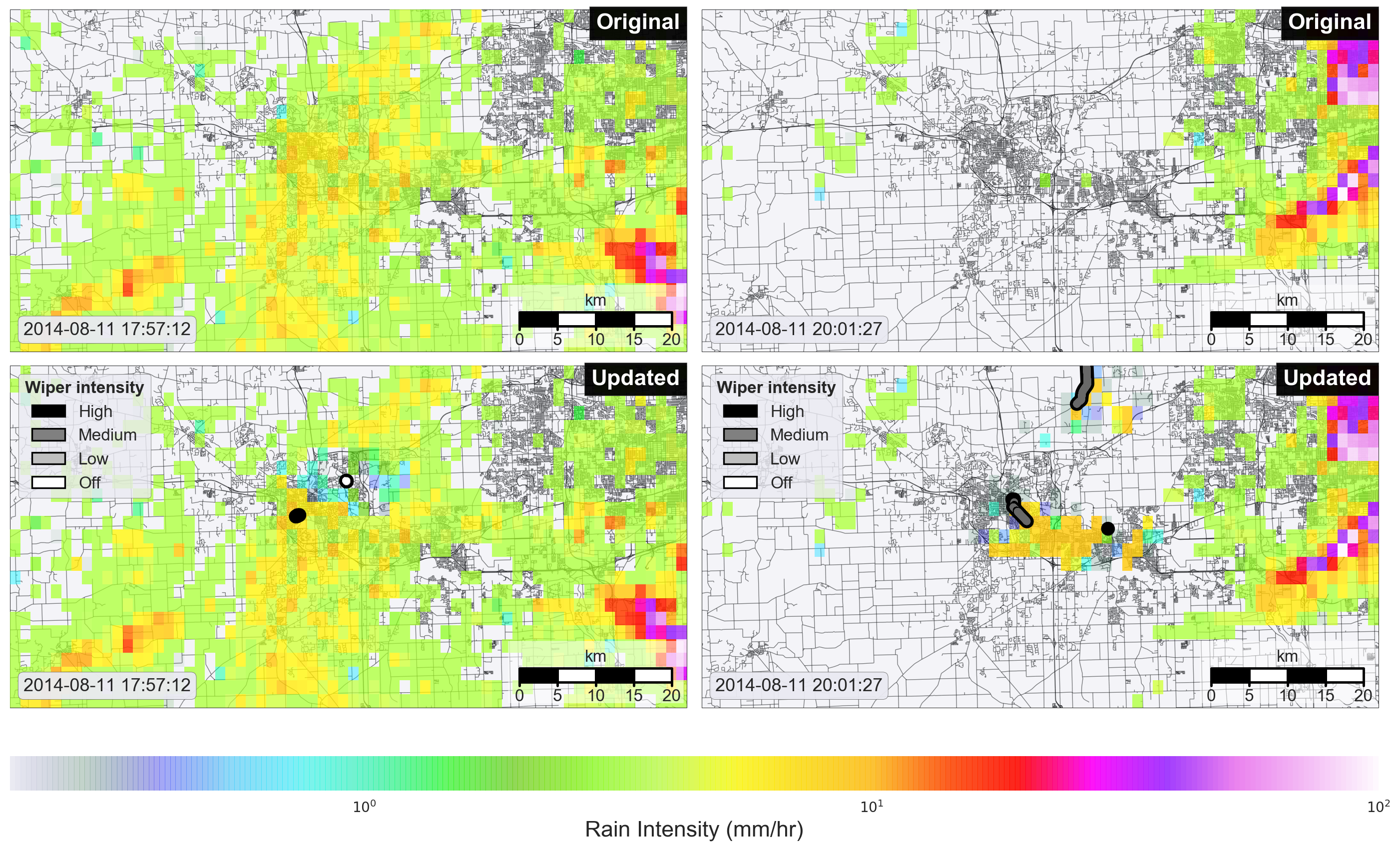}
\caption{\textbf{Original and updated rainfall maps}. Top (left and right):
  Original weather radar rainfall intensity map. Radial radar scans have been
  resampled to a 1 km grid to ensure computational tractability. Bottom (left
  and right): updated rainfall intensity map, combining radar data with wiper
  measurements using the Bayesian filter. In the bottom left panel, a ``hole''
  in the rainfall field occurs when a vehicle detects no rain in a location
  where radar alone estimated rain. In the bottom right panel, vehicles detect
  rainfall where radar previously did not detect rainfall.}
\label{updated_map}
\end{figure*}

Based on the observation that wiper measurements are a strong binary predictor
of rainfall, we develop a Bayesian filtering framework that combines radar
rainfall estimates with wiper observations to generate corrected rainfall maps.
Radar is used to estimate a prior distribution of rainfall intensities. This
prior is then updated with wiper observations to produce a corrected rainfall
field. The results of this filtering procedure are demonstrated in Figure
\ref{updated_map}, which shows the original rainfall field (top) along with the
corrected rainfall field (bottom). In some cases (left panel), vehicles detect
no rain in regions where radar had previously estimated rain. In these cases,
the updated product reduces the rainfall field in the proximity of the vehicle.
In other cases (right panel), the updated product predicts rainfall in regions
where little to no rainfall was observed in the original dataset. This outcome
shows that windshield wiper activity is sensitive to intermittent rainfall
events that radar may not otherwise detect. To see the full evolution of the
rainfall field under both the original and corrected data sets, refer to Video
S1.

\begin{figure}[!htb]
\centering
\includegraphics[width=4in]{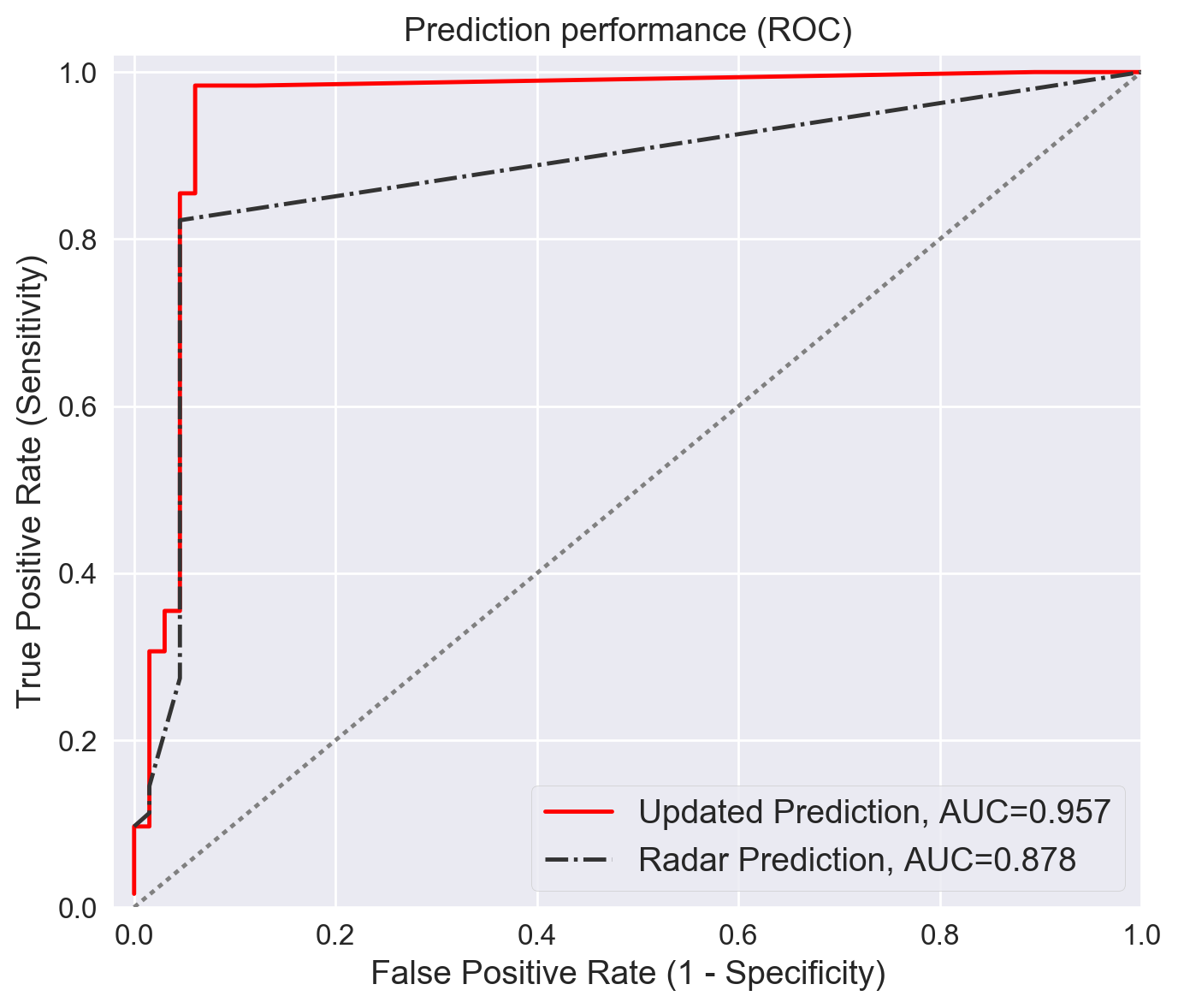}
\caption{\textbf{Binary classification performance of the updated rainfall
    product}. Receiver operator characteristic (ROC) curves indicate the
  rainfall state prediction accuracy for the original radar estimate and the
  updated (wiper-corrected) data product. The area under the curve (AUC)
  measures overall classification performance.}
\label{fig:roc_curve}
\end{figure}

The wiper-corrected rainfall field predicts the binary rainfall state with greater
accuracy than the radar-only data product. To validate the wiper-corrected rainfall
field, we use an iterated ``leave-one-out'' approach, in which an updated
rainfall field is generated while excluding a vehicle, and the resulting data
product is compared against the measured rainfall state of the omitted vehicle.
Repeating this process for each vehicle yields the receiver operator
characteristics shown in Figure \ref{fig:roc_curve}. These curves map the
relationship between the TPR and TNR for both the original rainfall field (radar
only) and the corrected rainfall field (radar and wiper). Curves located closer to
the upper-left corner (i.e. those with a larger area under the curve) exhibit
the best performance, given that they have a large true positive rate and a
small false negative rate. Based on these curves, it can be seen that the
corrected data product performs consistently better than the original radar
product at predicting the presence or absence of rain, with a TPR and TNR close
to unity. These results confirm that inclusion of vehicle-based measurements
enables improved prediction of the underlying rainfall field.

\section*{Discussion}

The enhanced rainfall maps developed in this study have the potential to assist
in the real-time operation of transportation and water infrastructure. In
particular, high accuracy rainfall field estimates will enable improved
prediction of flash floods in urban centers, and will help to inform real-time
control strategies for stormwater systems. As mentioned previously, flash flood
forecasting is contingent on high-resolution areal rainfall estimates, with
accurate measurements on the order of 500 m or finer required for forecasting in
urban areas. By enabling real-time validation and filtering of radar rainfall
estimates, vehicle-based sensors may help fill measurement gaps and improve the
prediction of flood events near roadways. Monitoring of roadways is especially
important given that in the US, roughly 74\% of flood fatalities are vehicle
related \cite{Doocy_2013}. While the vehicles used in this study provide only
binary measurements of rainfall state, many newer vehicles feature optical rain
sensors that are capable of measuring precipitation rate directly. When combined
with the Bayesian sensor fusion framework described in this study, these optical
rain sensors may enable robust mapping of rainfall volumes at the fine spatial
and temporal scales needed for high-accuracy flash flood forecasting. Moreover,
as connected and autonomous vehicles become more widely adopted, the spatial
coverage and measurement certainty of this new rainfall sensing modality will be
even further enhanced.

In addition to assisting with flash flood response, high-precision rainfall data
products may one day inform the operation of new ``smart'' water infrastructure.
Recent work has highlighted the potential of ``smart'' water systems to mitigate
water hazards through real-time control of distributed gates, valves and pumps
\cite{Bartos_2017, Kerkez_2016, Wong_2016, Wong_2016b, Mullapudi_2017}. When
informed by accurate and timely data, these systems can significantly reduce
operating costs, prevent combined sewer overflows, and halt the degradation of
aquatic ecosystems by adaptively reconfiguring water infrastructure in real time
\cite{Bartos_2017, Kerkez_2016}. However, recent findings suggest that optimal
control strategies for ``smart'' water systems are highly sensitive to the
location, timing and intensity of rainfall inputs \cite{Wong_2017}. In this
regard, the wiper-corrected rainfall product presented in this study may help to
enable more fine-grained control of water infrastructure by reducing uncertainty
in conventional rainfall field estimates.

\section*{Conclusions}

This study generates enhanced probabilistic rainfall maps by combining
conventional radar-based precipitation fields with ubiquitous windshield wiper
measurements from almost 70 unique vehicles. We find that while windshield wiper
intensity is a poor predictor of rainfall intensity, wiper activity is a
stronger predictor of binary rainfall state than conventional radar and
gage-based data sources. With this result in mind, we develop a novel Bayesian
filtering framework that combines a radar-based rainfall prior with binary
windshield wiper observations to produce an updated rainfall map. We find that
the Bayesian filtering process is effective at detecting changes in the rainfall
field that conventional measurement technologies may otherwise miss. We validate
the updated rainfall data product by assessing its ability to reproduce the
binary rainfall state anticipated by an omitted vehicle. Based on this analysis,
we find that the corrected rainfall field is better at predicting the binary
rainfall state than the original radar product. As connected vehicles become
more widespread, the ubiquitous sensing approach proposed by this study may one
day help to inform real-time warning and control systems for water
infrastructure by providing fine-grained estimates of the rainfall field.

\section*{Materials and Methods}

\subsection*{Evaluating vehicle-based measurements}
In the first part of this study, we assess the degree to which windshield wiper
activity serves as a proxy for both rainfall intensity and binary rainfall
state. First, wiper measurements are compared against conventional rainfall
measurement technologies to determine if there is a direct relationship between
wiper intensity and rain intensity. Next, we assess the degree to which each
data source reflects the ground truth rainfall state by comparing measurements
from all three sources (gages, radar and wipers) with vehicle-based video
footage. Video footage provides instantaneous visual confirmation of the
rainfall state (raining or not raining), and is thus taken to represent the
ground truth. We characterize the binary classification performance of each
technology in terms of its true positive and true negative rates. 

To ensure that our analysis is computationally tractable, we isolate the study
to a subset of three storms in 2014. We assess the validity of our procedure for
storms of different magnitudes by selecting a large storm (2014-08-11), a
medium-sized storm (2014-06-28) and a small storm (2014-06-12). Storms are
selected during the summertime months to avoid conflating rainfall measurements
with snow measurements. The year 2014 is chosen because it is the year for which
the greatest number of vehicles are available. Unless otherwise specified, data
are co-located using a nearest neighbor search. For comparison of wiper and gage
readings, we select only those gages within a 2 km range of any given vehicle.

\subsection*{Data sources}
We consider four data sources: (i) stationary rain gages, (ii) weather
surveillance radar, (iii) vehicle windshield wiper data, and (iv) vehicle
dashboard camera footage. We provide a brief description of each data source
here:

\begin{description}
\item [Gage data] are obtained from personal weather stations maintained by the
  Weather Underground \cite{wunderground}. Within the city of Ann Arbor (Michigan),
  Weather Underground hosts 21 personal weather stations, each of which yield
  rainfall estimates at a time interval of approximately 5 minutes. Locations of
  gages are indicated by blue circles in Figure \ref{overview_map}. Although
  verified gage data from the National Weather Service (NWS) and the National
  Oceanic and Atmospheric Administration (NOAA) are available, Weather
  Underground gages are selected because (i) NOAA and NWS each maintain only a
  single gage in the city of Ann Arbor, meaning that intra-urban spatial
  variations in precipitation intensity cannot be captured, and (ii) the
  temporal resolution of NOAA and NWS gages are relatively coarse for real-time
  applications (with NOAA offering a maximum temporal resolution of 15 minutes
  and NWS offering a maximum temporal resolution of 1 hour).

\item [Weather radar observations] are obtained from NOAA's NEXRAD Level 3 Radar
  product archive \cite{nexrad3}. We use the ``Instantaneous Precipitation
  Rate'' data product (p176). Radar precipitation estimates are obtained at a
  temporal resolution of 5 minutes, and a spatial resolution of 0.25 km by 0.5
  degree (azimuth). Radar station KDTX in Detroit is used because it is the
  closest radar station to the City of Ann Arbor. Radial radar scans are
  interpolated to cartesian coordinates using a nearest neighbor approach.

\item [Vehicle-based wiper intensities] are obtained from the University of
  Michigan Transportation Research Institute (UMTRI) Safety Pilot Model
  Deployment database \cite{umtri}. For each vehicle, this dataset includes time
  series of latitude, longitude, and windshield wiper intensity at a temporal
  resolution of 2 milliseconds. Windshield wiper intensity is given on an
  ordinal scale from 0 to 3, with 0 indicating that the wiper is turned off, 1
  representing the lowest wiper intensity, and 3 representing the highest wiper
  intensity. A wiper reading of 4 indicates that the vehicle's ``mister'' is
  activated, distinguishing between wiper use for rain removal and wiper use for
  windshield cleaning. For the year 2014, 69 unique vehicles are available in
  the UMTRI dataset. However, typically less than ten vehicles are active at any
  given time during the observation period. Vehicles with no sensor output or
  invalid readings were removed from the dataset prior to the analysis (see the
  Supplementary Note for more details).

\item [Camera observations] are also obtained from the UMTRI vehicle database
  \cite{umtri}. Located on the inside of each vehicle, cameras provide streaming
  video footage of the windshield, side-facing windows, rear-facing windows, and
  the driver. For the purposes of validation, we use the front-facing windshield
  camera. Camera frames are manually inspected for rain drops striking the
  windshield. Time intervals where rain is observed are classified as
  ``raining''; similarly time intervals where no new droplets are observed are
  classified as ``not raining''. Manual inspection and labeling of the video
  data was performed independently by two reviewers to ensure robustness.
\end{description}

\subsection*{A Bayesian filtering framework}

In the second part of this study, we develop a Bayesian filtering framework that
combines binary wiper observations with radar-based rainfall intensity
measurements to generate corrected rainfall maps. In simple terms, the Bayesian
filter generates an updated rainfall field map, in which binary (on/off) wiper
measurements adaptively correct the underlying radar rainfall field. Windshield
wiper status is taken to represent a measurement of the ground truth binary
rainfall state, given that it is a better predictor of the binary rainfall state
than radar- or gage-based measurements. Under this framework, four distinct
cases are possible. If both the wiper and radar measure precipitation, the radar
reading is taken to be correct, and the original rainfall field remains the
same. Similarly, if neither the wiper nor the radar measure precipitation, the
radar rainfall field remains zero. However, if the radar measures precipitation
at a target location and the wiper does not, then the filter will update the
rainfall field such that rain intensity is reduced within the proximity of the
vehicle (with a decay pattern corresponding to the Gaussian kernel and an
intensity of zero at the location of the wiper reading). Similarly, if the wiper
measures precipitation, but the radar measures no precipitation, the rainfall
intensity will be increased around the proximity of the vehicle (with an
intensity defined by the empirical intensity distribution associated with the
given wiper intensity).

A more formal description of the filtering framework is given here in terms of a
noisy sensor model (for additional details, see \cite{Park:2018aa}). Consider a
noisy sensor model in which each sensor produces a binary measurement given a
target state. The target state is represented as a random tuple
$\bm{z}=(q,\bm{I})$ where $q$ is a location state (e.g. the latitude and
longitude at the target), and $\bm{I}$ is an information state (e.g. the
precipitation intensity at the target) with all the random quantities indicated
by bold italics. We denote by $M_t$ the event that sensors correctly measure the
intensity, and by $\overline{M}_t$ the event that sensors fail to measure the
intensity correctly. The joint measurement likelihood at any time $t$ is given
by:

\begin{equation}
  p(M_t | \bm{z},x_t)
  \label{eq:prob1}
\end{equation}

where $x_t$ represents the locations of the sensors at time $t$. Equation
\ref{eq:prob1} yields the probability distribution of precipitation intensity
measurement at $q$ by sensors at $x_t$. The expected value of Equation
\ref{eq:prob1} with respect to $\bm{I}$ is equivalent to the rainfall intensity
experienced at the location $q$. Because the effective range of the wipers is
limited, we account for the probability of detection as a function of the
distance between the sensor and the target. We denote by $D_t$ the event that
sensors detect the target, and by $\overline{D}_t$ the event that sensors fail
to detect the target at time $t$. The probability of detecting a target located
at $q$ by sensors located at $x_t$, $p(D_t | q,x_t)$, is taken to decay with
increasing distance to the sensor. Using the law of total probability, the
conditional probability of a correct measurement is then given by:

\begin{align}
    p(M_t | \bm{z},x_t)=p(M_t | \bm{z},D_t,x_t) p(D_t|q,x_t) + p(M_t | \bm{z},\overline{D}_t,x_t)p(\overline{D}_t |  q,x_t)
\end{align}

where $D_t$ is conditionally independent of $\bm{I}$ when conditioned on $q$.
For example, consider $x_t = (0,0)$, and $q = (q_1,q_2)$. If the decay function
is taken to be a 2D Gaussian centered at $x_t$ with covariance matrix $\sigma
\mathbf{I}$ where $\mathbf{I}$ is a 2 by 2 identity matrix, then:

\begin{equation}
  p(D_t | q,x_t) = 
  \widetilde{\eta}_t
  \frac{1}{2 \pi \sigma^2} \exp\left(-\frac{q_1^2 + q_2^2 }{2 \sigma^2}\right)
\end{equation}

Where $\widetilde{\eta}_t$ is a normalization constraint. If the target is
\emph{not} detected (i.e., $\overline{D}_t$), then the measurement is assumed to
be unreliable, and the likelihood, $p(M_t|\bm{z},\overline{D}_t,x_t)$, is
modeled using a prior distribution. If there is no prior information available,
the function is modeled using a uniform distribution. Now let $b_t(\bm{z})$
represent the posterior probability of the precipitation intensity given a
target location $q$ at time $t$. Using Bayes’ Theorem, $b_t(\bm{z})$ can be
formulated:

\begin{equation}
 b_t(\bm{z}) = \eta_t p(M_t|\bm{z},x_t) b_{t-1} (\bm{z}),\,\,\,\,\,\,\,\,\,\,t=1,2,\dots 
\end{equation}

Where $\eta_t$ is a normalization constant and $b_{0}$ is uniform if no information is available at $t=0$. This filtering equation forms the
basis of the rainfall field updating algorithm. To reduce computational
complexity, the filtering operation is implemented using a Sequential Importance
Resampling (SIR) Particle Filter \cite{berzuini1997dynamic}.

The results of the Bayesian sensor fusion procedure are evaluated by determining
the proportion of instances where the combined data product is able to predict
the binary rainfall state. We characterize the true and false positive rates for
the largest storm event (2014-08-11) using an iterated ``leave-one-out''
cross-validation approach. First, a single vehicle is removed from the set of
vehicles. The Bayesian update procedure is then executed using all vehicles
except the excluded vehicle, and an updated rainfall map is generated. Next, the
rainfall states predicted by the corrected rainfall field (radar and wiper) and
the original rainfall field (radar only) are compared against the rainfall
states predicted by the omitted vehicle. The performance of each data product is
evaluated based on its ability to reproduce the binary rainfall state observed
by the omitted vehicle. Performing this process iteratively yields the true and
false positive rates for both the original (radar only) and updated (radar and
wiper) rainfall fields. This procedure is repeated for each vehicle in the set
of vehicles to generate Receiver-Operator Characteristic (ROC) curves, which
characterize the true and false positive rates across an ensemble of
simulations.

\bibliography{sr}

\section*{Acknowledgements}
Funding for this project was provided by MCubed (grant 985), the Ford Motor
Company--University of Michigan Alliance (grant N022977), and the University of
Michigan. Vehicle metadata and camera footage are provided courtesy of the
University of Michigan Transportation Research Institute. 

\section*{Author contributions statement}
M.B. wrote the paper, performed the analysis, and helped with the implementation
of the filtering algorithm. H.P. developed, implemented, and validated the
filtering algorithm. T.Z. analyzed the dashboard camera data and assisted with
analysis of the windshield wiper data. B.K. and R.V. originated the concept of
the study, guided the development of the methods, and assisted in writing the
paper. Additional inspection and labeling of vehicle dashboard footage was
performed by Aditya Prakash Singh. All authors reviewed the manuscript.

\section*{Additional information}

\subsection*{Data access links}
Upon publication, code and data for this study will be made available at:
\texttt{github.com/kLabUM/vehicles-as-sensors}.

\subsection*{Competing financial interests statement}
The authors declare no competing interests.

\clearpage

\section*{Supplementary Materials}

\subsection*{Supplementary note on binary detection performance}

Binary detection performance is sensitive to a number of factors, including
the temporal resolution of the ground truth data and the configuration of wiper
sensors. While these factors can affect the magnitude of binary classification
performance, under all scenarios considered, wiper measurements are a better
detector of the binary rainfall state than either radar or gage measurements. 

Binary detection performance can be affected by the temporal resolution at which
the ground truth data is collected. To ensure robustness, labeling of
vehicle footage was performed independently by two reviewers. The first reviewer
labeled the observed rainfall state for each vehicle over all three days of the
study period (2014-06-12, 2014-06-28, 2014-08-11) at a temporal interval of 1
minute. A second reviewer labeled the observed rainfall and wiper state for the
largest storm event (2014-08-11) at an enhanced time resolution of roughly 3
seconds. Due to the time-intensive nature of labeling video data at this
temporal resolution, and due to the strong agreement between the two labeled
datasets, this second round of labeling was not performed for the remaining two
days (2014-06-12 and 2014-06-28). Despite the difference in time resolution,
manual labeling of the video data showed strong agreement. Taking the
high-resolution dataset to represent the ground truth rainfall state (and
aggregating the high-resolution dataset to the temporal resolution of the
low-resolution dataset), the true positive rate of the low temporal-resolution
camera observations was 92.6\%, while the true negative rate was 99.3\%.
Agreement in terms of positive detection was lower due to the difference in
temporal resolution between the two sources. The low-resolution camera
observation dataset classifies each minute-long interval as either ``raining'' or
``not raining''. However, the high-resolution ground truth dataset contains many
instances in which part of a given minute-long interval contains rain, and part
does not. Thus, when the high-resolution dataset is aggregated to match the
resolution of the low-resolution dataset, there are more intervals where some
amount of rain is detected (yielding more instances of positive detection
overall). A similar mismatch occurs if the low-resolution dataset is
interpolated to match the time resolution of the high-resolution dataset. This
time resolution mismatch also affects comparisons between the ground truth and
other data sources (e.g. wiper, radar and gages). In general, the difference in
classification performance between data sources decreases when the ground truth
dataset is aggregated in time. Differences in classification performance become
more pronounced when a high-resolution ground truth dataset is used.

Many vehicles exhibited data quality issues such as non-reporting wiper sensors,
malfunctioning wiper sensors, or unobservable wiper modes. These data quality
issues may impact the performance of the wiper as a classifier, but are largely
attributable to the fact that the data is taken from a pilot study in which
sensor configurations are not standardized. For some vehicles, wiper sensors
were simply not configured to report wiper data. In these instances, the
reported wiper value was zero for the entire observation period even though
wiper movement was observed during manual inspection of the dashboard footage.
Vehicles for which wiper sensors were not configured were removed from the
analysis. Other vehicles exhibited malfunctioning or poorly configured sensors.
For instance, in some cases the wiper intensity fluctuated between 0 and 1 at a
frequency on the order of milliseconds---a behavior which is clearly not
possible for a human driver. Video footage confirmed that the sensor was
malfunctioning during these time periods. Malfunctioning vehicles were also
removed from the analysis. Perhaps the most common data quality issue, however,
is that several vehicles exhibited unobservable wiper modes. In this case,
sensors were configured to report some wiper intensity states but not others.
For example, the sensor may report the wiper intensity when the wiper switch is
in a ``continuous'' mode, but may not report the wiper intensity when the wiper
is placed in a manual ``wipe'' mode. These cases could only be detected by
manual inspection of the camera footage. These data issues can largely be
attributed to the fact that the sensor data is taken from a pilot study in which
sensor configurations vary from vehicle to vehicle. As manufacturers standardize
sensor configurations for connected vehicles, the relevance of these issues is
likely to diminish.

The performance of the wiper as a classifier can be improved by (i) comparing
wiper data against a ground truth dataset obtained at a high temporal
resolution, and (ii) correcting errors in the wiper sensor readings. When manual
observations of the wiper state are used to correct unobservable wiper modes,
and the resulting corrected wiper data is compared to the 3-second resolution
camera observations, the binary classification performance over weather radar is
significantly enhanced: the true positive rate of the wiper data is 5.2\% higher
than radar, while the true negative rate is 7.7\% higher. Table S1 shows the
true and false positive rates for all technologies (during the 2014-08-11 storm
event) when these two conditions are met.

\clearpage

\subsection*{Fig. S1}

\begin{figure}[!htb]
\centering
\includegraphics[width=\textwidth]{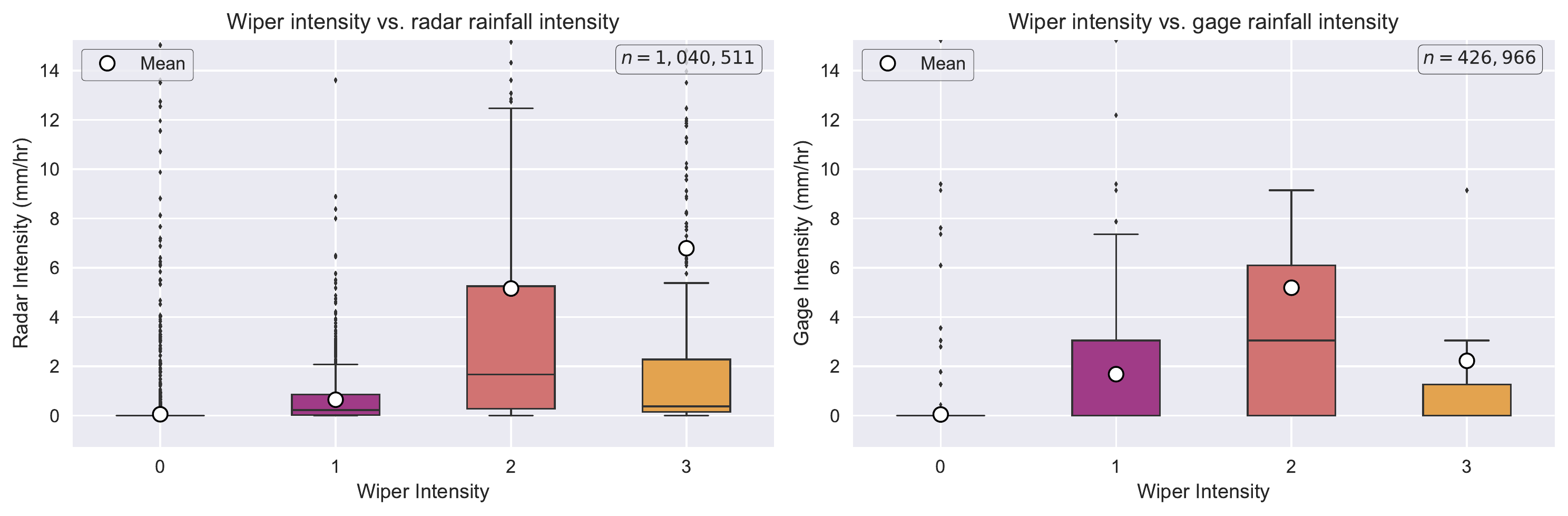}
\caption{\textbf{Comparison of radar, gage and wiper intensities for three storm events
  on 6/12/2014, 6/28/2014, and 8/11/2014.} The left panel shows the distribution
  of radar precipitation measurements associated with each wiper intensity. The
  right panel shows the distribution of gage precipitation measurements
  associated with each wiper intensity for vehicles located within 2 kilometers
  of the gage. Note that the range limitation reduces the number of data points
  available. No clear relationship is observed between wiper intensity and
  rainfall intensity.}
\end{figure}

\clearpage

\subsection*{Table S1}

\begin{table}[!htb]
  \centering
  \begin{tabular}{| l | l l l |}
    \hline
    Metric & Gage & Radar & Wiper \\
    \hline
    True Positive Rate (\%) & 55.1 & 91.8 & 97.0 \\
    True Negative Rate (\%) & 96.9 & 87.4 & 95.1 \\
    \hline
  \end{tabular}
  \caption{\textbf{Classification performance of each rainfall measurement
      technology when using high-temporal resolution ground truth data, and
      correcting misreporting wiper states}. These binary performance metrics hold
    when (i) ground truth observations at a resolution of 2.4 seconds are used,
    and (ii) manual corrections are made to the wiper state according to the
    wiper state in the observed camera footage (i.e. unobservable wiper modes
    are corrected).}
  \label{tab:tab1}
\end{table}

\clearpage

\subsection*{Video S1}

  Original rainfall field (top) vs. updated data product (bottom) for a
  large storm event on 2014-08-11. Vehicle paths can be seen in the bottom
  frame, with windshield wiper intensities indicated by greyscale intensity from
  off (white) to high intensity (dark grey).

\end{document}